\documentclass{eds2019}

\def\be{\begin{equation}}
\def\ee{\end{equation}}
\def\bea{\begin{eqnarray}}
\def\eea{\end{eqnarray}}

\newcommand{\Photo}{\includegraphics[height=35mm]{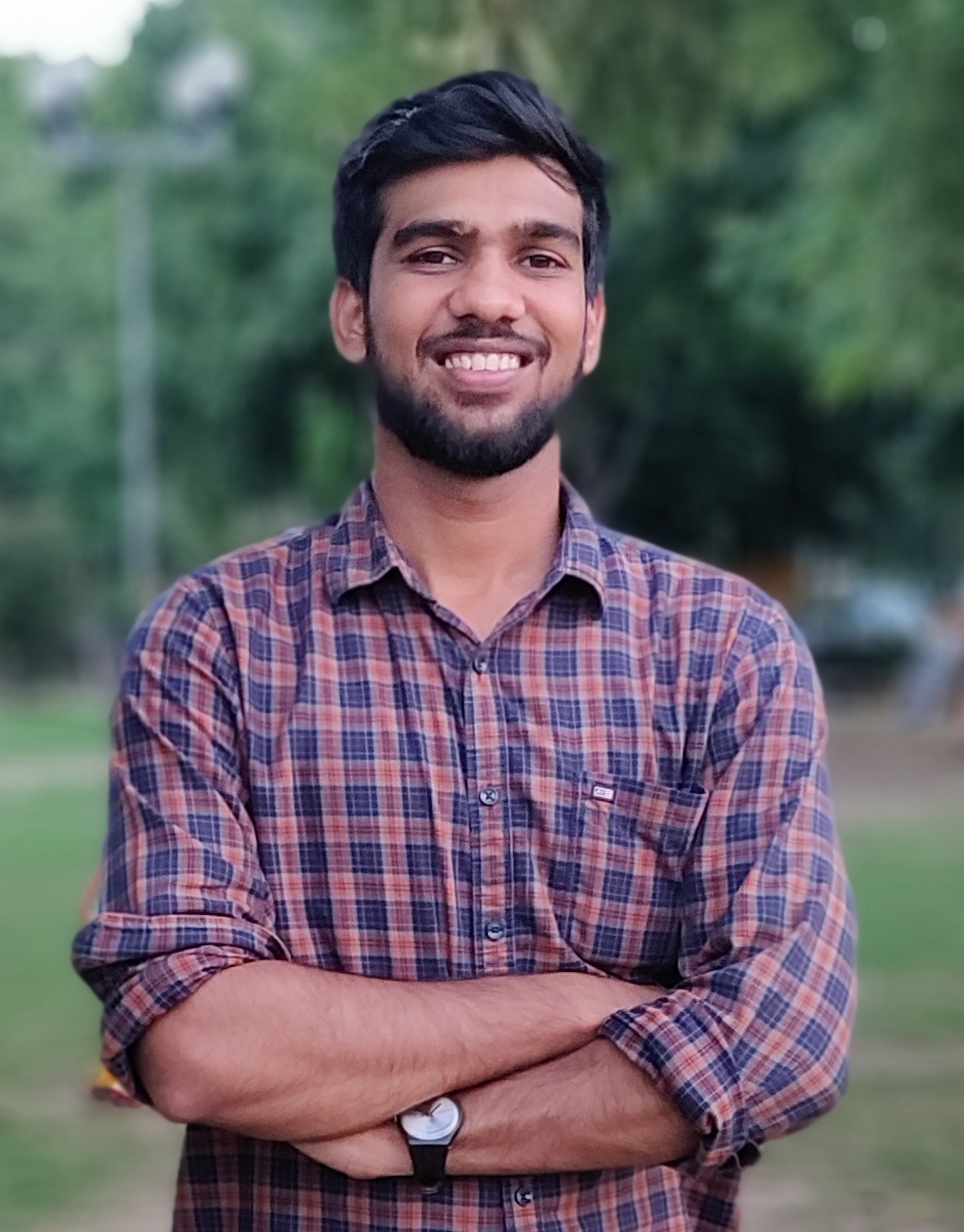}}

\begin{document}
\vspace*{4cm}
\title{Latest results on Soft QCD and DPS from the CMS experiment}

\author{ Rajat Gupta on behalf of the CMS Collaboration}

\address{Panjab University, Chandigarh}

\maketitle\abstracts{
We present latest results of soft and small-x QCD measurements with the CMS experiment, such as minimum bias/underlying event physics, and studies on forward jet production.}

\section{Introduction}

Quantum chromodynamics (QCD), the theory of strong interactions, is undoubtedly a very rich and successful theory. There are interesting phenomena in strong interactions which have yet to be seen, phenomena that are ultimately related to the properties of color confinement and asymptotic freedom of the strong interactions. These effects can take place in special corners of phase space accessible at the Large Hadron Collider (LHC). More concretely, it is not clear whether a gluon-gluon recombination mechanism takes place at low values of the fraction of the nucleon momentum \textit{x} carried by its partonic constituents. Said mechanism is believed to slow down the rapid growth of the nucleon's structure function at very small values of \textit{x}. 
	On the other hand, we have to refine our understanding of the underlying dynamics in low momentum exchange processes in hadronic collisions. The description of these effects rely on phenomenological models, whose parameters are tuned based on fits to data. Dedicated measurements provide valuable inputs for Monte Carlo event generators, which are of great importance for precision measurements of Standard Model processes and searches for New Physics at the LHC. 
	In this context, we briefly discuss the following recent results by the CMS experiment:

\begin{itemize}
\item Measurement of inclusive very forward jet cross sections in proton-lead collisions at $\sqrt{s} =$ 5.02 TeV~\cite{One}.
\item Measurement of the average very forward energy as a function of the track multiplicity at central rapidities in proton-proton collisions at 13 TeV~\cite{Two}.
\item Measurement of charged particle spectra in minimum-bias events from proton-proton collisions at $\sqrt{s} =$ 13 TeV~\cite{Three}.
\item Measurement of the underlying event in $t\overline{t}$ dilepton events at 13 TeV~\cite{Four}.
\item Evidence for WW production from double-parton interactions in proton-proton collisions at $\sqrt{s} =$ 13 TeV~\cite{Five}.
\end{itemize}

\section{Measurement of inclusive very forward jet cross sections in proton-lead collisions at $\sqrt{s}_{NN} =$ 5.02 TeV}

\begin{figure}[htbp]
\centering
\includegraphics[width=0.45\textwidth]{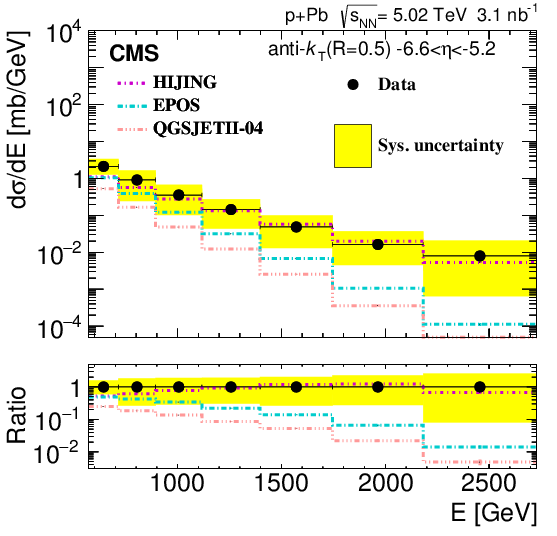}
\includegraphics[width=0.45\textwidth]{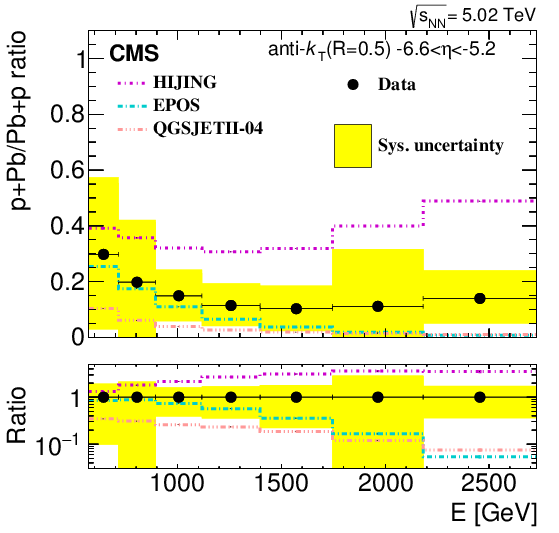}\\
 \caption{Stable-particle-level differential jet cross section as a function of jet energy in proton- lead collisions at 5.02 TeV. The Pb+p measurement is depicted left, and the ratio of the p+Pb to Pb+p cross sections is displayed right~\protect\cite{One}} 
\label{fig:jet}
\end{figure}

Measurements of the differential inclusive forward jet cross sections in proton-lead collisions at $\sqrt{s}_{NN} = $5.02 TeV have been presented. The measurements are performed in the laboratory pseudorapidity range -6.6 $< \eta <$ -5.2, and as a function of jet energy. Jet production in hadron-hadron interactions is useful tool to study the parton structure of hadrons. Collisions with either the incoming proton (p+Pb) or the incoming ion (Pb+p) directed towards the negative $\eta$ hemisphere are studied. The jet cross sections are unfolded to stable-particle level cross sections with $p_{T} >$  3 GeV and compared to predictions from various Monte Carlo event generators as shown in Figure~\ref{fig:jet}. The cross section ratio for p+Pb to Pb+p data as a function of jet energy has also been measured, and exhibits a much smaller systematic uncertainty than the individual spectra. The so-far unexplored kinematic phase space covered by this measurement is sensitive to the parton densities and their evolution at low fractional momenta. Models incorporating various implementations of gluon saturation have been confronted with data. No model is, however, currently able to describe all aspects of the data.

\section{Measurement of the average very forward energy as a function of the track multiplicity at central pseudorapidities in proton-proton collisions at $\sqrt{s} =$ 13 TeV}

The average energy per event in the pseudorapidity region -6.6 $< \eta <$ -5.2 was measured as a function of the observed central track multiplicity ($|\eta| <$ 2) in proton-proton collisions at a centre-of-mass energy of 13 TeV. The measurement is presented in terms of the total energy as well as its electromagnetic and hadronic components. The very forward region covered by the data contains the highest energy densities studied in proton-proton collisions at the LHC so far. This makes the present data relevant for improving the modelling of multiparticle production in event generators of ultra-high energy cosmic ray air showers. 

\begin{figure}[htbp]
\centering
 \includegraphics[width=0.48\textwidth]{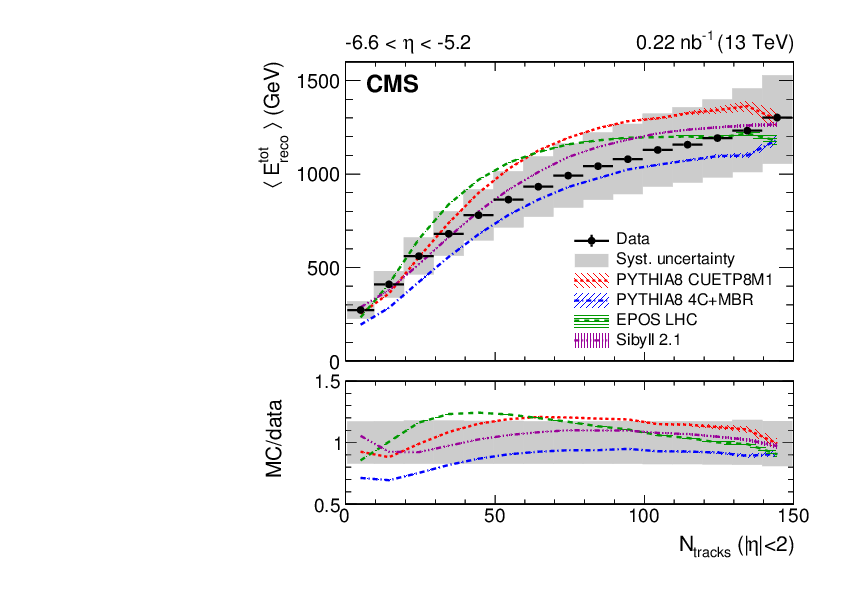}\label{fig:forward-1}
\includegraphics[width=0.48\textwidth]{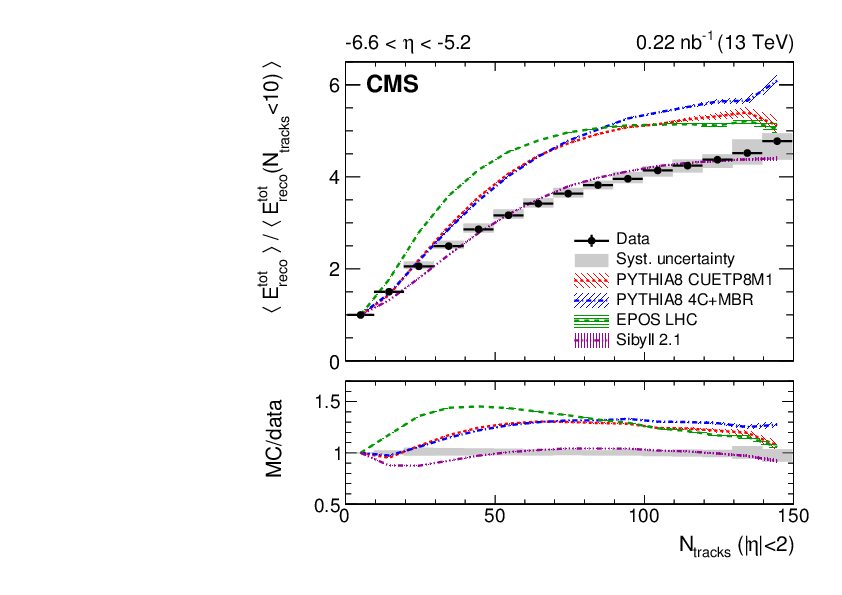}\label{fig:forward-2}
 \includegraphics[width=0.48\textwidth]{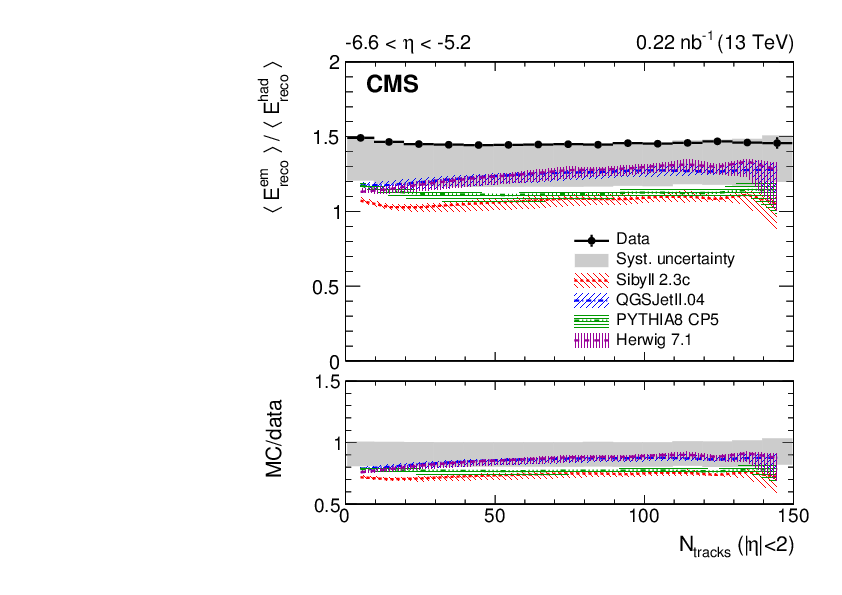}\label{fig:forward-3}\\
 \caption{Top Left:  Average total energy reconstructed in the CASTOR calorimeter as a function of the number of reconstructed tracks for $|\eta| <$ 2. Top Right: Average total energy reconstructed in the CASTOR calorimeter normalised to that in the first bin ($N_{ch} <$ 10) as a function of the number of reconstructed tracks for $|\eta| <$ 2. Bottom: Ratio of average electromagnetic and hadronic energies reconstructed in the CASTOR calorimeter as a function of the number of reconstructed tracks for $|\eta| <$ 2~\protect\cite{Two}.} 
\label{fig:forward}
\end{figure}

As shown in top left of Figure~\ref{fig:forward}, the measured average total energy as a function of the track multiplicity is described by all models with reasonable discrepancies. This demonstrates that the underlying event parameter tunes determined at central rapidity can be safely extrapolated to the very forward region within experimental uncertainties. A shape analysis indicates, however, that there are significant differences among the models and large deviations from the data. The generator SIBYLL 2.1 gives the best description of the measured multiplicity dependence of the average total energy as shown in top right of Figure~\ref{fig:forward}.

The ratio between the electromagnetic and hadronic energies is also presented in bottom row of Figure~\ref{fig:forward}. The data exhibit a larger fraction of electromagnetic energy than the models, and disagree with the two most recent model tunes, i.e. SIBYLL 2.3c and PYTHIA 8 CP5. Therefore, these models cannot explain the muon deficit in ultra-high energy air shower simulations since the data indicate that even more energy must be channelled into the electromagnetic part of the cascade and is thus lost for the generation of further hadrons.

\section{Measurement of charged particle spectra in minimum-bias events from proton-proton collisions at $\sqrt{s} =$ 13 TeV}
\begin{figure}[htbp]
\centering
\includegraphics[width=0.42\textwidth]{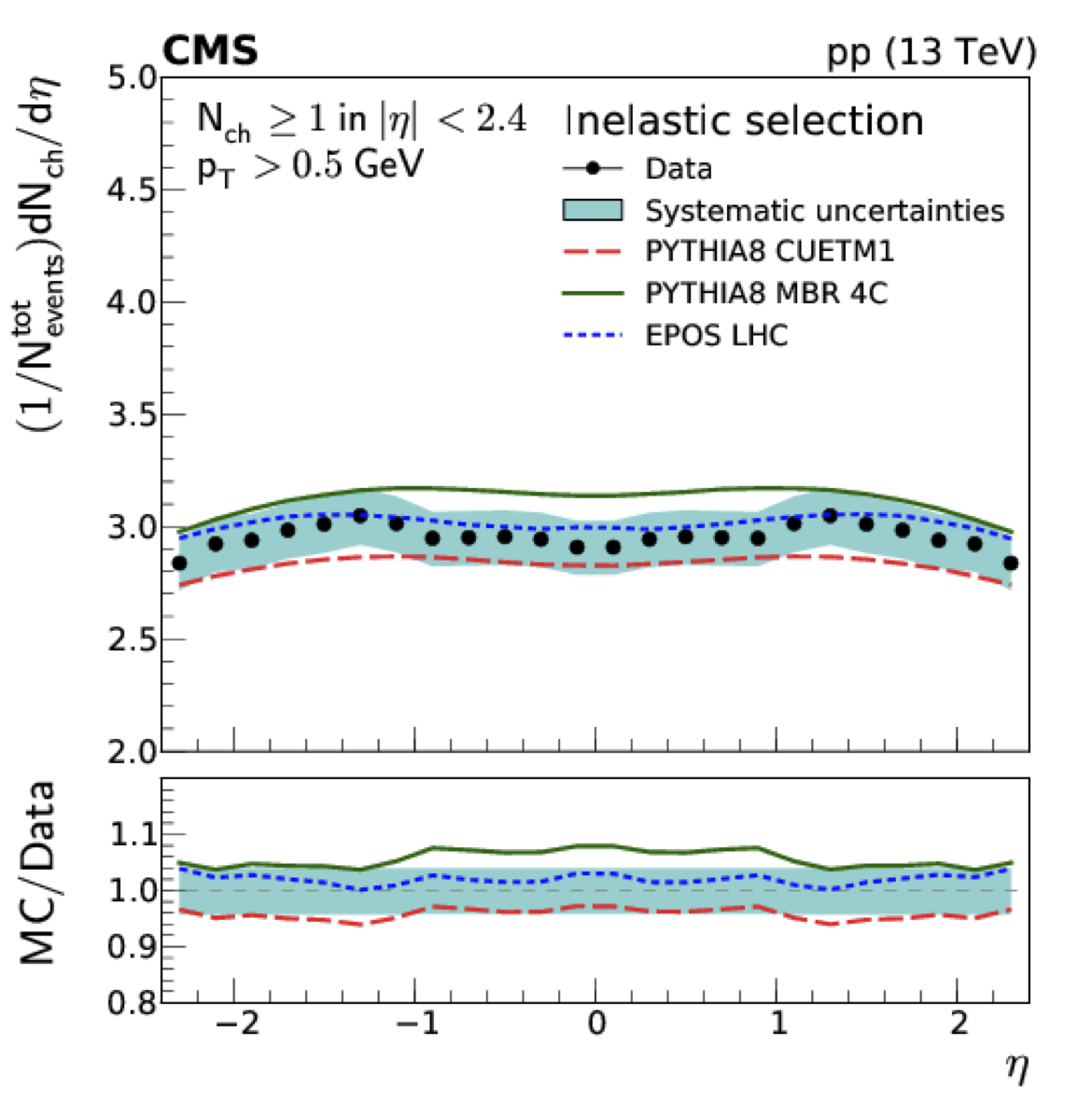}\label{fig:MB-a}
 \includegraphics[width=0.42\textwidth]{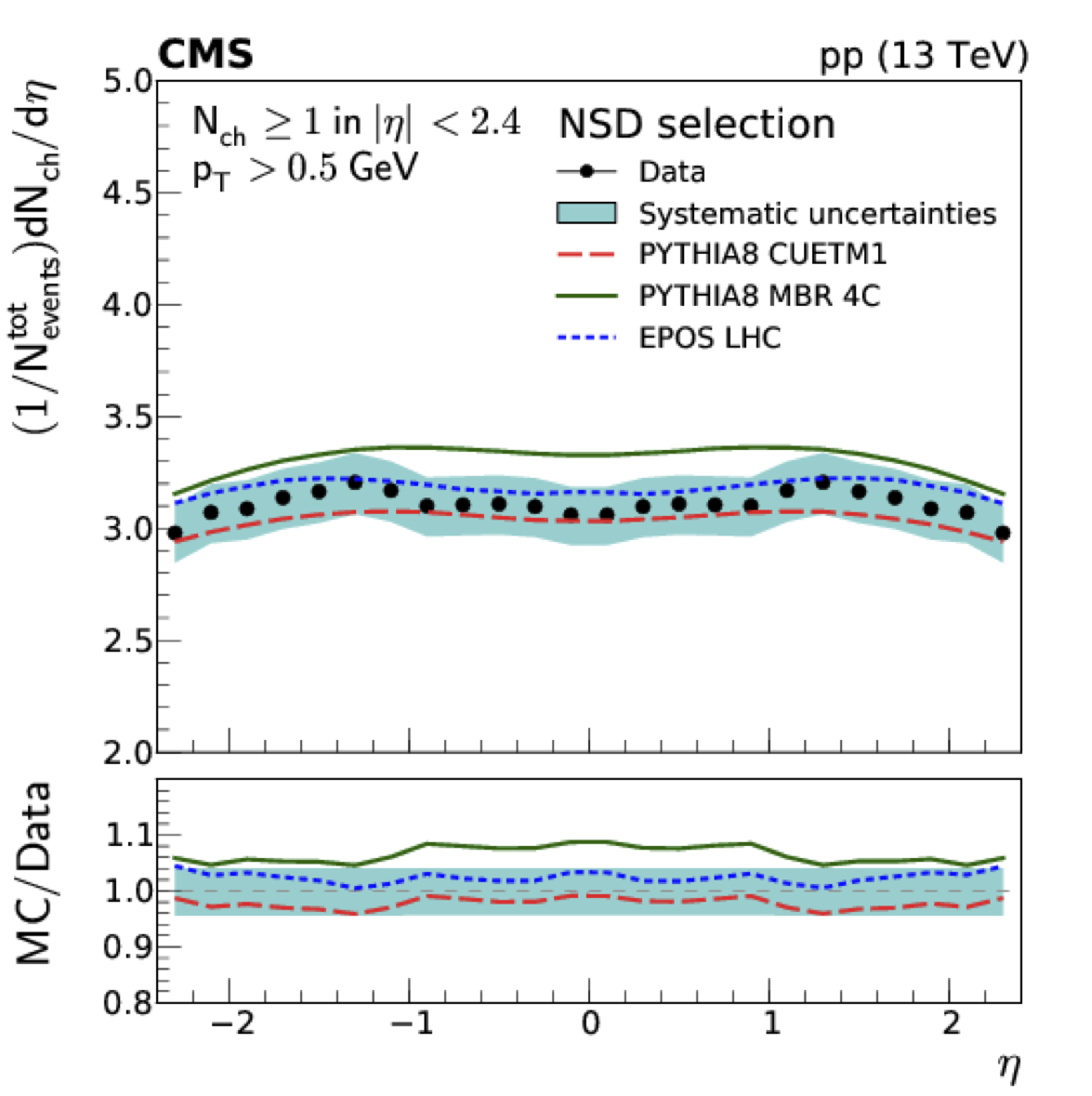}\label{fig:MB-b}\\
\includegraphics[width=0.42\textwidth]{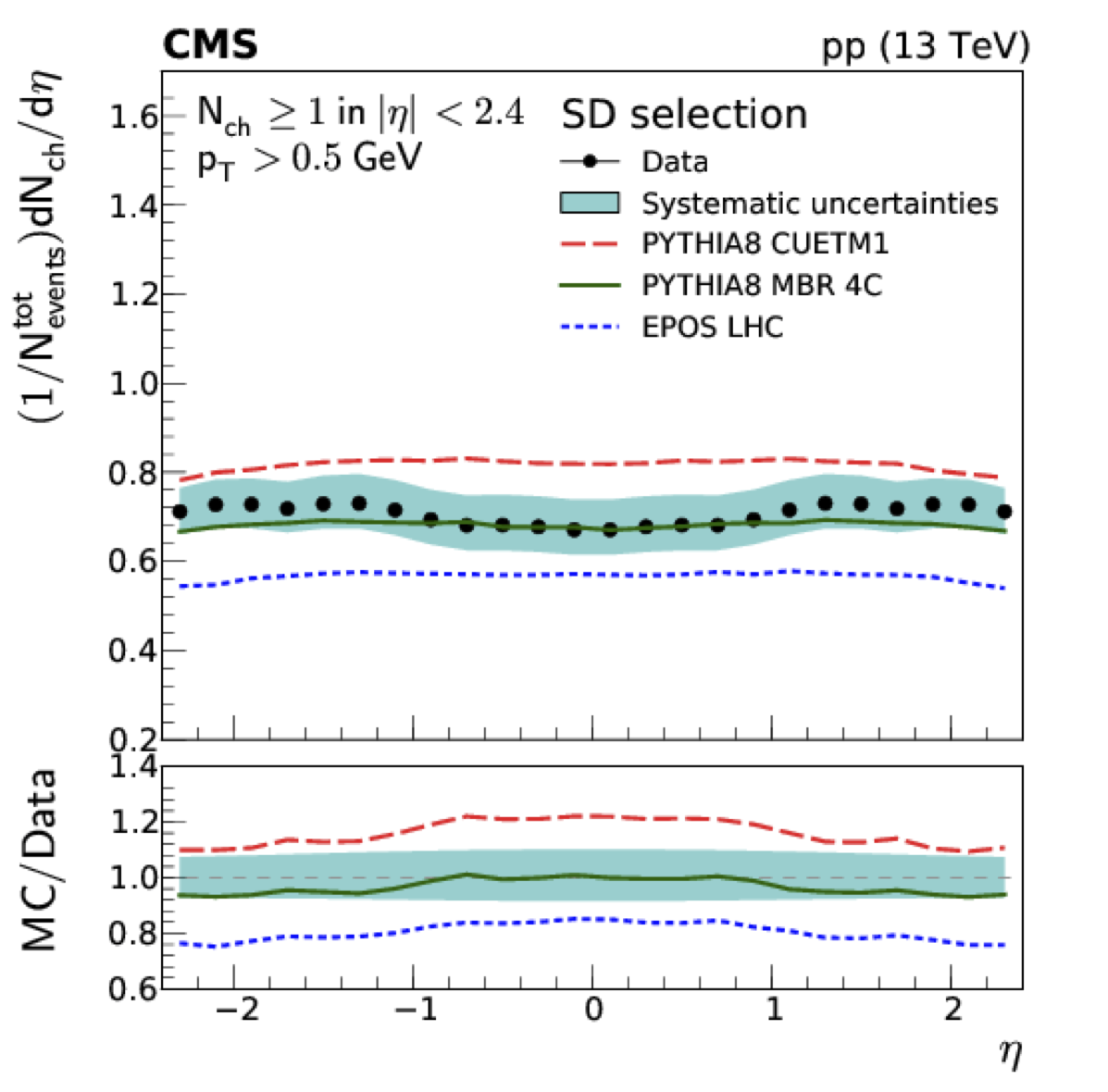}\label{fig:MB-c}
\includegraphics[width=0.42\textwidth]{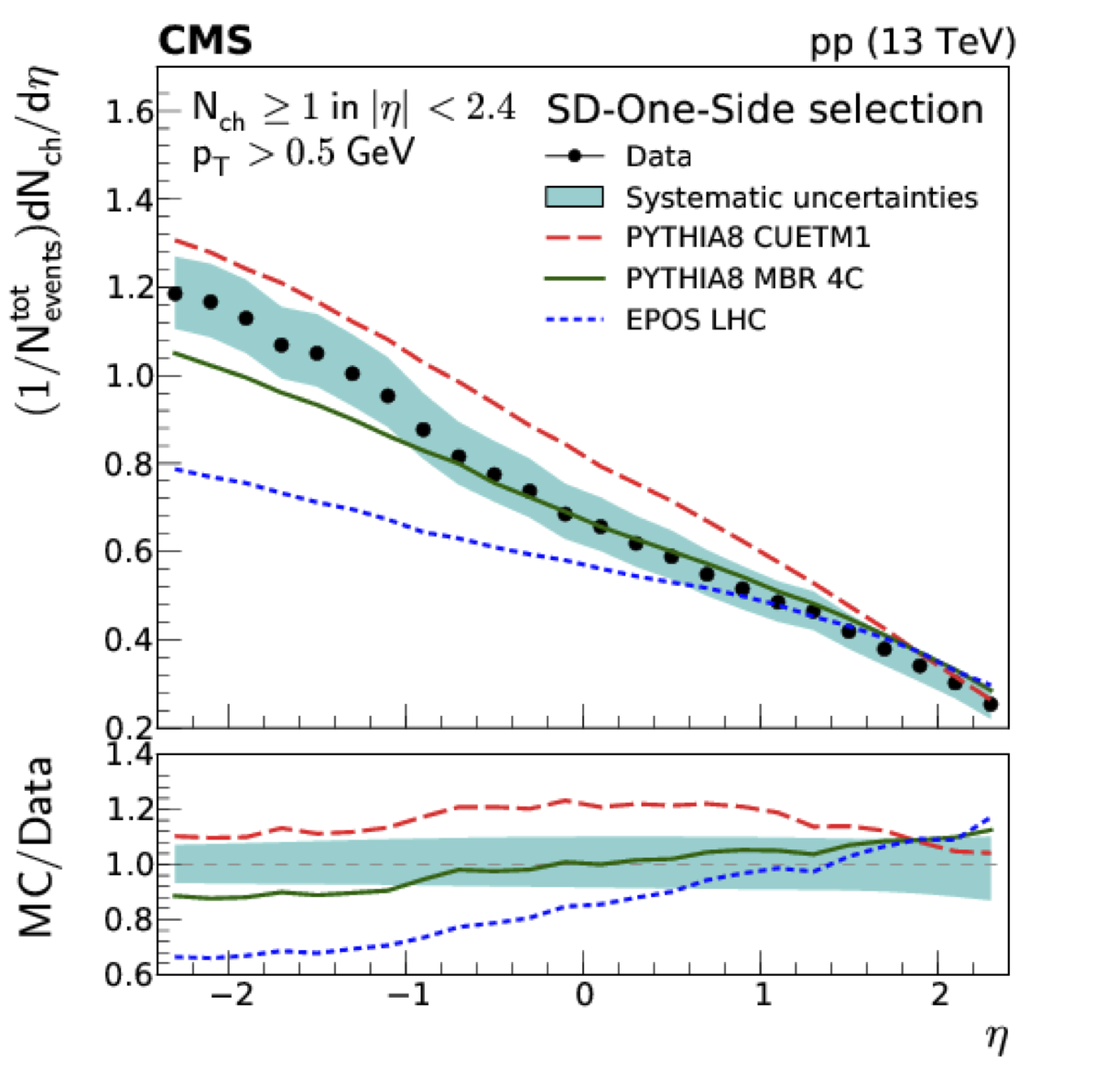}\label{fig:MB-d}\\
 \caption{Differential cross section distribution of leading and subleading jet's transverse momentum and pseudorapidity for data and standard model predictions for signal and various background processes. Ratio plot at the bottom represents the ratio of data over total MC simulated events~\protect\cite{Three}.} 
\label{fig:MB}
\end{figure}

Particle production without any selection bias arising from the requirement of the presence of a hard scattering process is known as minimum bias (MB). The bulk of these events occur at low momentum exchanges between the interacting partons inside the hadrons, where diffractive scattering or multiple partonic interactions (MPI) play a significant role. 
One can characterize MB events by means of charged particle distributions. Charged particle distributions are measured for charged particles with $p_{T} >$ 0.5 GeV and $|\eta| <$ 2.4 for events collected with the CMS MB trigger. The measured distributions are presented for different event data samples based on the calorimeter activity in the forward region by requiring the presence of at least one tower with energy above 5 GeV in the acceptance region 3 $< |\eta| <$ 5, and in some cases with a veto condition for towers less than a given threshold value. The different event classes are as non-single diffractive enriched sample (NSD-enhanced) when there is calorimeter activity in both sides, as single diffractive enriched (SD-enhanced) when there is calorimeter activity on one side and a veto on the opposite side, and as inelastic when there is calorimeter activity on at least one side of CMS. The distribution labelled as SD-One-Side enhanced sample corresponds to the symmetrized distribution constructed from the SD-minus and SD-plus enhanced samples. 
The normalized particle distribution is measured as a function of the charged particle pseudorapidity for the three different selections (see Figure~\ref{fig:MB}). The results are unfolded to particle level. PYTHIA8 CUETM1, PYTHIA8 MBR 4C and EPOS LHC results are compared to the data. PYTHIA8 MBR 4C describes reasonably well the data for the SD-enhanced samples, but overestimates the yield in central pseudorapidities for the non-diffractive samples. PYTHIA8 CUETM1 and EPOS LHC give a fair description for the non-diffractive samples, but they are off w.r.t. data for the SD-enhanced selection. 

\section{Underlying Event Measurements}

The underlying events (UE) surround the hard scattering of a hadronic interaction and can receive contributions from initial and final state radiation (ISR/FSR), QCD evolution or color reconnection (CR) and additional partonic interactions in the same collision. Measuring the UE is crucial for a proper understanding of the interaction and for MC model tuning.



\begin{figure}[htbp]
\centering
\includegraphics[width=0.45\textwidth,height=160pt]{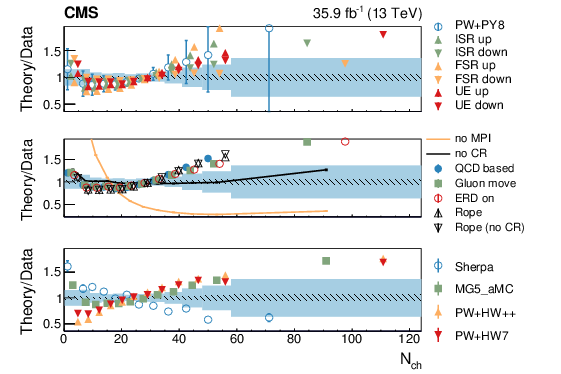}\label{fig:UE-1}
\includegraphics[width=0.45\textwidth]{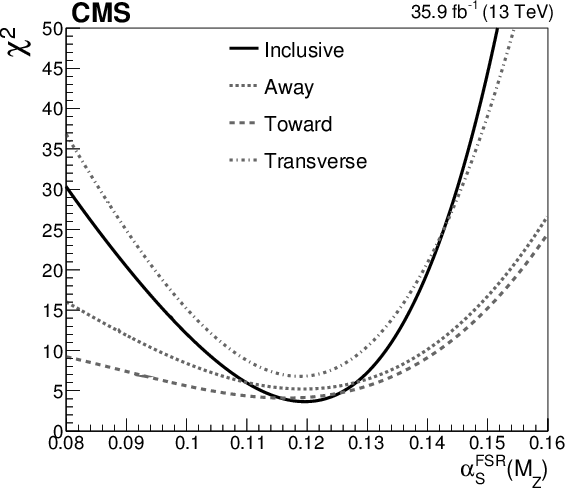}\label{fig:UE-2}\\
 \caption{The normalized differential cross section as a function of Nch compared to predictions of different models (left), scan of the  as a function of the value of  employed in PW+P8 simulation~\protect\cite{Four}.} 
\label{fig:UE}
\end{figure}

The study of the UE in $t\overline{t}$ events provides a direct test of its universality at higher energy scales than those probed in minimum bias or DY events. This is relevant as a direct probe of color reconnection (CR), which is needed to confine the initial QCD color charge of the t quark into color-neutral states. 
	Various characteristics, such as the multiplicity of the selected charged particles, the flux of momentum, and the topology or shape of the event have different sensitivity to the modeling of the recoil, the contribution from MPI and CR, and other parameters, and hence used to characterize UE activity. Figure~\ref{fig:UE} (left) shows the normalized differential cross section measured as function of $N_{ch}$. The majority of the distributions analyzed indicate a fair agreement between the data and the POWHEG + PYTHIA8 setup with the CUETP8M2T4 tune, but disfavour the setups in which MPI and CR are switched off, or in which $\alpha_{s}^{FSR} (M_{z})$ is increased. The data also disfavour the default configurations in HERWIG++, HERWIG7, and SHERPA. It has been furthermore verified that, as expected, the choice of the next-to-leading order matrix-element generator does not impact significantly the expected characteristics of the UE by comparing predictions from POWHEG and MADGRAPH5\_MC@NLO, both interfaced with PYTHIA8. The sensitivity of these results to the choice of $\alpha_{s}^{FSR} (M_{z})$ in the parton shower is tested by performing a scan of the $\chi^{2}$ value as a function of  $\alpha_{s}^{ISR} (M_{z})$ or  $\alpha_{s}^{FSR} (M_{z})$ as shown in Figure~\ref{fig:UE} (right).  A value of  $\alpha_{s}^{FSR} (M_{z})$ is obtained, which is lower than the one assumed in the Monash tune and used in the CUETP8M2T4 tune. The new CMS default tune CP5 uses $\alpha_{s}^{FSR} (M_{z}) =  \alpha_{s}^{FSR} (M_{z}) =$ 0.118 [13].

\section{Double Parton Scattering Measurements}

\begin{figure}[htbp]
\centering
\includegraphics[width=0.45\textwidth]{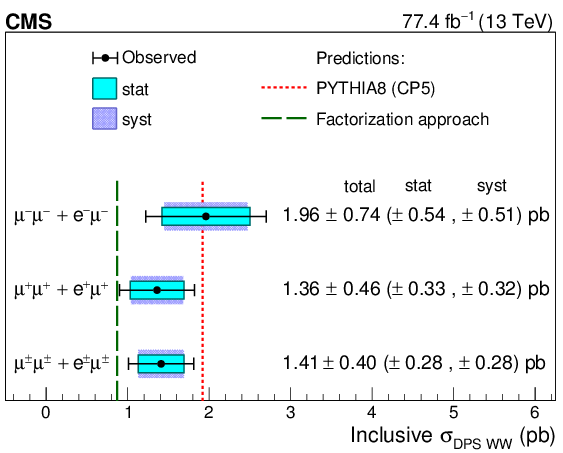}\label{fig:DPS-2}
\includegraphics[width=0.45\textwidth]{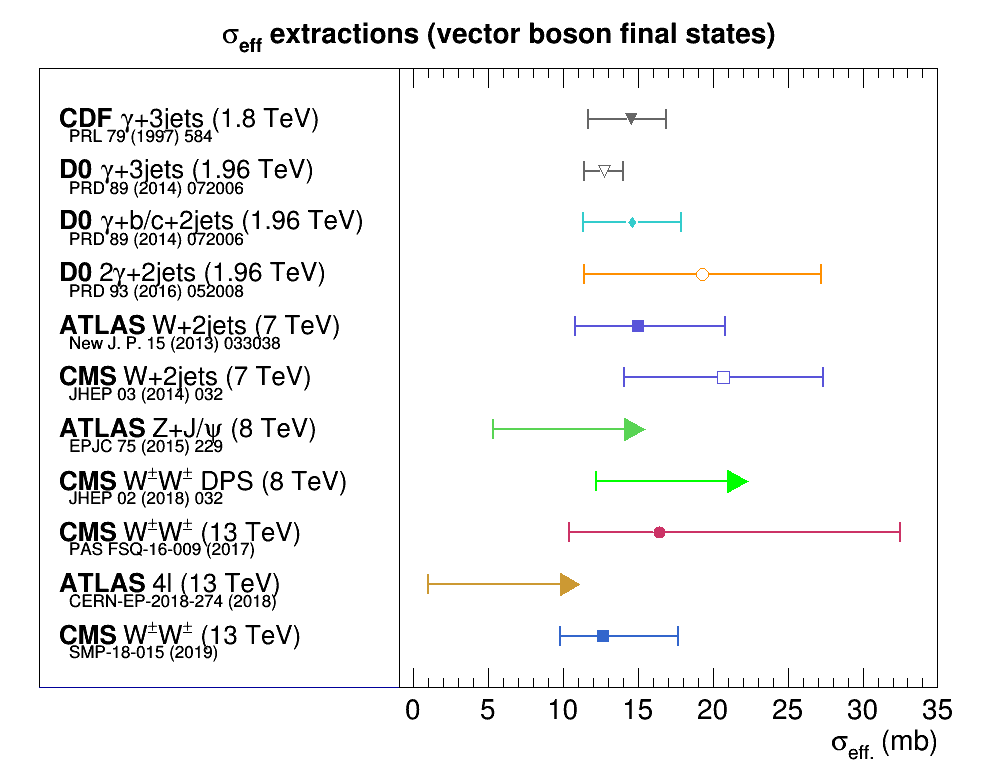}\label{fig:DPS-3}\\
 \caption{Left: Observed cross section values for inclusive DPS WW production from the two lepton charge configurations and their combination. These values are obtained from the extrapola- tion of the observed DPS W±W± cross section to the inclusive WW case. Right: comparison of extracted  using various processes at the previous experiments and CMS measurements~\protect\cite{Five}} 
\label{fig:DPS}
\end{figure}

The study of DPS processes provides valuable information on the transverse distribution of partons in the proton and on the parton correlations in the hadronic wave function. A study of WW production from double-parton scattering (DPS) processes in proton-proton collisions at $\sqrt{s} =$ 13 TeV has been reported. The analyzed data set corresponds to an integrated luminosity of 77.4 $fb^{-1}$, collected using the CMS detector in 2016 and 2017 at the LHC. The WW candidates are selected in same-charge electron-muon or dimuon events with moderate missing transverse momentum and low jet multiplicity. Multivariate classifiers based on boosted decision trees are used to discriminate between the signal and the dominant background processes. A maximum likelihood fit is performed to extract the signal cross section, which is compared to the predictions from simulation and from an approximate factorization approach as shown in Figure~\ref{fig:DPS}. A measurement of the DPS WW cross section is achieved for the first time, and a cross section of 1.41 $\pm$ 0.28 (stat) $\pm$ 0.28 (syst) pb is extracted with an observed significance of 3.9 standard deviations. This cross section leads to an effective cross section parameter of $12.7_{-2.9}^{+5.0}$ mb. This result is the first evidence for DPS WW production.

\end{document}